\documentclass[letterpaper]{article}

\usepackage{geometry}
\usepackage{url}
\usepackage{graphicx}
\usepackage{color}
\usepackage{amssymb}
\usepackage{amsmath}
\newcommand{\omitit}[1]{} 

\usepackage{xcolor,colortbl}
\definecolor{Gray}{gray}{0.85}

\let\footnote=\endnote
\title{A Survey of Game Theoretic Approaches for\\
 Adversarial Machine Learning in Cybersecurity Tasks}
\author{Prithviraj Dasgupta, Joseph B. Collins\\[0.1in]
Distributed Systems Section (Code 5583)\\
Information Management \& Decision Architectures (IMDA) Branch\\
Information Technology Division\\
U.S. Naval Research Laboratory, Washington D.C., USA\\[0.2in]
E-mail: \{raj.dasgupta, joseph.collins\}@nrl.navy.mil
}

\date{}

\begin{document}
\maketitle


\begin{abstract}
Machine learning techniques are currently used extensively for automating various cybersecurity tasks. Most of these techniques utilize supervised learning algorithms that rely on training the algorithm to classify incoming data into different categories, using data encountered in the relevant domain. A critical vulnerability of these algorithms is that they are susceptible to adversarial attacks where a malicious entity called an adversary deliberately alters the training data to misguide the learning algorithm into making classification errors. Adversarial attacks could render the learning algorithm unsuitable to use and leave critical systems vulnerable to cybersecurity attacks. Our paper provides a detailed survey of the state-of-the-art techniques that are used to make a machine learning algorithm robust against adversarial attacks using the computational framework of game theory. We  also discuss open problems and challenges and possible directions for further research that would make deep machine learning-based systems more robust and reliable for cybersecurity tasks.
\end{abstract}

{\bf Keywords:} Machine learning, adversarial learning, game theory.

\section{Introduction}
Adversarial learning ~\cite{Tygar11} is an instance of machine learning where two entities called the learner and adversary attempt to learn a prediction mechanism for data related to a problem domain at hand, albeit with different objectives. The learner's objective in learning the prediction mechanism is to correctly predict or classify the data. In contrast, the adversary's objective is to imperceptibly coerce the learner into making incorrect predictions for the data in the future. A very popular instance of adversarial learning is email spam filtering~\cite{Tygar11,Huang11}. Here, the learner is the spam filter with a prediction mechanism that classifies incoming email into two categories, spam or non-spam. On the other hand, the adversary is the spammer who in addition to generating spam email, also tries to add, remove or alter certain words or characters in the email text~\cite{Dalvi04} so that it can disguise non-spam email as spam, and vice-versa. If the spammer is succcessful, the spam filter ends up mis-classifying the altered non-spam emails as spam (false positives) or the altered spam emails as non-spam (false negatives). Both these mis-classifications could be dangerous for the integrity of the email filtering system, not only do they block legitimate email and allow potentially malicious spam email to pass through, but they also reduce the confidence in the email classification performed by the spam filter. Adversarial learning poses a severe cybersecurity threat in several domains that employ machine learning-based classifier systems including automated email spam filters and anti-virus software, image classification algorithms in defense and medical applications, and text-based sentiment analysis algorithms used on social media data. To combat these challenges, researchers have proposed several techniques that aim to make the learner's classifier robust against adversarial attacks~\cite{Huang11}. Many of these techniques employ a popular framework of decision making at the intersection of mathematics, economics and computer science called game theory~\cite{Fudenberg91,Myerson97,LeytonBrown09}. Game theory provides an attractive tool for adversarial learning as it provides a means to mathematically model the learner and adversary's behaviors in terms of defense and attack strategies and determine suitable strategies to reduce the learner's loss from adversarial attacks. In this survey, we focus on such game theory-based techniques that have been used to make machine learning algorithms robust against adversarial attacks.


\begin{figure*}[thb!]
\begin{center}
\includegraphics[width=2.0in]{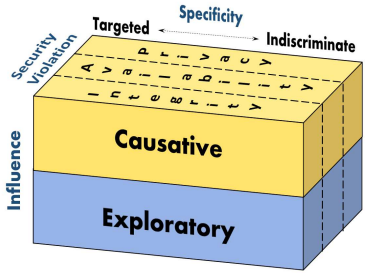}
\caption{{\small{Taxonomy of different types of adversarial attacks along three dimensions proposed in~\cite{Huang11}. Influence and security violation-based attacks are divided into distinct categories marked by solid and dashed dividing lines. Specificity attacks range over a continuous spectrum marked as dotted arrows.}}}
\label{fig_adv_taxonomy}
\end{center}
\end{figure*}

The rest of the paper is structured as follows: in the next section we provide background information on adversarial learning and game theory. Following that, we summarize the contributions of game theory-based adversarial learning approaches and solution techniques. Then, we discuss open issues and challenges for future research directions in game theory-based modeling of adversarial learning, and, finally we conclude. Also, in the rest of the paper, in accordance with machine learning literature, we assume that the output of the learner's prediction mechanism classifies the data into a finite set of classes, each class is identified with an output label. Finally, for legibility, while following most of the existing literature in this area, we assume that the learner uses a classifier for its prediction mechanism. In general, the learner could use any other prediction mechanism like clustering, ranking or regression.

\section{Background: Adversarial Learning and Games}
\label{sec_background}
\begin{figure*}[thb!]
\begin{center}
\includegraphics[width=4.8in]{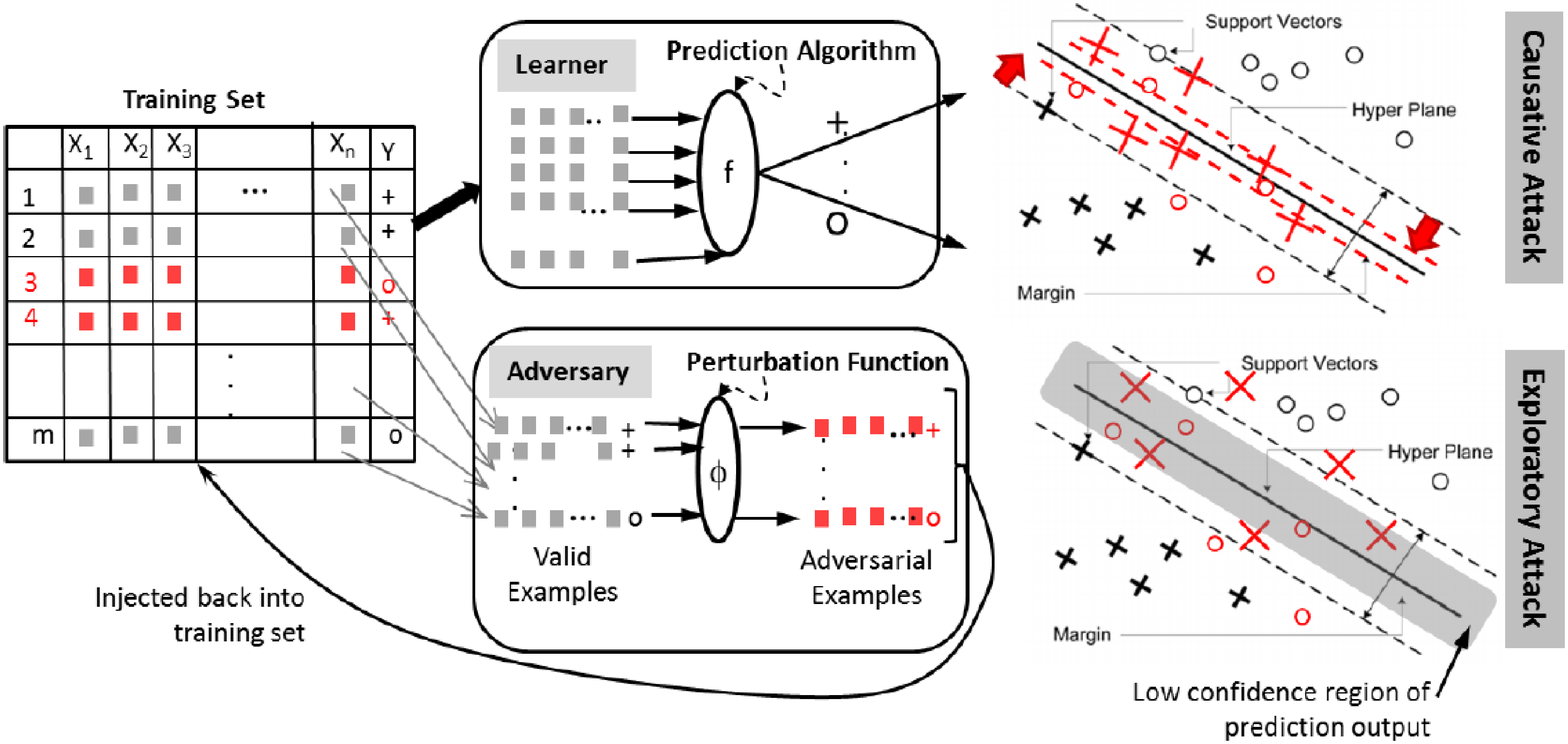}
\caption{{\small{Adversarial learning scenario showing the two types of influence-based adversarial attacks - causative (top right) and exploratory (bottom right). The learner uses a support vector machine to classify input. Red color input represents adversarial examples that are created by the adversary using a perturbation function $\phi()$ from valid examples from the training set and re-injected into the training set.}}}\textsl{•}
\label{fig_adv_attacks}
\end{center}
\end{figure*}

{\em Adversarial Learning.} Adversarial learning deals with techniques used by a machine learning-based prediction mechanism such as a classifier to make itself robust against adversarial attacks. In~\cite{Huang11}, authors provide a comprehensive overview of adversarial learning techniques. Their work includes a taxonomy for adversarial learning while characterizing it along three dimensions - influence, specificity and security violation, as shown in Fig.~\ref{fig_adv_taxonomy}. Influence is the most relevant and widely researched dimension for adversarial learning because it characterizes the adversary based on its behavior, with the objective of developing appropriate learner strategies to counter the adversary's behavior. The influence dimension specifies two types of adversarial attacks called {\em causative} and {\em exploratory} (a.k.a. probing) that are illustrated in Fig.~\ref{fig_adv_attacks}. In {\em causative attacks} the adversary acquires data that was used to train the learner's classifier and modifies this data. This modified data, called adversarial data, is then used by the learner during further training of its classifier. This results in the learner learning an incorrect classifier that gives classification errors (false positives and false negatives) while testing or while using the classifier. In {\em exploratory attacks}, the adversary observes the output of the learner's classifier for different data and tries to discover its vulnerabilities (e.g., what data it mis-classifies). It then creates adversarial data that exploits those vulnerabilities to increase the classifier's mis-classification rate during test or application time. Two other dimensions of adversarial learning shown in Fig.~\ref{fig_adv_taxonomy} are security violation and specificity. The reader is referred to~\cite{Huang11}, for approaches that counter adversarial attacks along these three dimensions. The rest of this paper focuses specifically on how the influence dimension's causative and exploratory attacks are countered with game theory-based techniques.

{\em Non-cooperative Game.} Adversarial learning has been extensively modeled as a $2$-player, non-cooperative game. A non-cooperative game can be informally defined as an interaction between two or more players over a resource that has to be shared between the players. The game, represented in normal form, is given by $(N, A, U)$, where $N$ is the set of players, $A=\{A_i\}$ where $A_i$ is the set of actions for player $i$, and, $U=\{U_i\}$, were $U_i(a_i, a_{-i})$ gives a real-valued number called utility received by each player $i \in N$ when it selects action $a_i \in A_i$ while other players jointly select $a_{-i} \in A_{-i}, A_{-i} = \times A_{j \neq i}$. The utility of each player gives its preference over the different outcomes of the game resulting from different joint actions by the players. Player $i$'s strategy set $S_i=\{\pi(A_i): \pi(A_i)\geq 0\;\forall i, \sum_{A_i} \pi(A_i)=1\}$ specifies a probability distribution over its actions $A_i$. The outcome of a game is a strategy selected by each player. One of the most widely used techniques to calculate a player's strategy in a game is given by Nash equilibrium (NE). NE assumes that players behave rationally and each player $i$ plays its {\em best response} strategy given by $s_i$ that satisfies \(U_i(s_i, s_{-i}) \geq U_i(s'_i, s_{-i}), \forall i \in N\). The NE of a $2$-player game can be represented as a linear complimentarity problem and solved using linear programming or as a search problem~\cite{LeytonBrown09}. Below, we briefly mention a few aspects of games that are relevant to adversarial learning. Because adversarial learning is a $2$-player game between the learner and the adversary, we denote the two players using subscripts $l$ and $a$ respectively.

{\em Zero-sum vs. Non-zero Sum Game}. In a $2$-player, zero-sum game, the utilities of players, the learner and adversary, sum to zero. In other words, the gain in utility of the learner comes at the cost of loss of adversary's utilities and vice-versa. In a $2$-player, zero-sum game, the NE can be calulated using the minimax theorem, which says that the game's Nash equilibrium outcome is the same as its minimax outcome. The minmax outcome can be represented as constrained optimization problem and solved as a linear program. Adversarial learning has been extensively modeled as a $2$-player, zero sum game. However, recently Bruckner {\em et al.}~\cite{Bruckner11} observe that assuming adversarial learning is a zero-sum game is overly pessimistic - the utility loss of the learner might not equal the utility gain of the adversary. Consequently, they model adversarial learning as a non-zero sum or general sum game. Unfortunately, minmax theorem's result does not hold for general sum games, and, more complicated, general Nash equilibrium solution techniques~\cite{LeytonBrown09} have to be used to determine the learner and adversary's selected strategies.

{\em Simultaneous Move vs. Sequential Game.} In a simultaneous move game, players make their strategy selection simultaneously and cannot observe each others' strategy before selecting their own. In contrast, in a sequential move game, players take turns in selecting their strategies (or, making their moves). Adversarial learning has been modeled as the latter - the learner is the player making the first move or the leader, because it usually publishes its classifier and is not aware of the presence of the adversary~\cite{Huang11,Groshans13}. The adversary, on the other hand, is the follower because it can observe the learner's classifier and then make its move of selecting a suitable strategy to generate adversarial instances. Sequential move games are easier to solve than simultaneous move games because the follower knows the leader's selected strategy and can use this information to select its utility-maximizing strategy. The leader's strategy selection technique, however, does not have information about the follower's selected strategy. Consequently, while selecting its strategy, the leader has to incorporate this uncertainty about the follower's strategy using a type of game called a Bayesian game, as described next.

{\em Bayesian Game}. The normal form game mentioned above assumes that each player has information about the utilities of other players, for each action. This assumption might not be valid in many practical, real-life scenarios because it is unrealistic for a player to have accurate information about competing player's utilities. For example, in adversarial learning, the learner might not have accurate information of the adversary's cost to generate adversarial data~\cite{Bruckner11,Groshans13} or the adversary might not have accurate information about the learner's classification cost~\cite{Lowd05}. The problem is addressed through a Bayesian game~\cite{Harsanyi68}, where each player is assumed to have a set of types. The utility that a player gets from each of its actions now also depends on its types. A player does not know the exact type of other players, but it knows the probability distribution over the types. Based on this information, a player can calculate expected utilities, conditioned on the other players' types. Some researchers have modeled adversarial learning as a Bayesian sequential-move game, where the learner assumes a set of types for the adversary along with probability distribution over the types. It then selects a strategy based on its expected utilities conditioned on the prior probabilities of the different adversary types, as discussed in the section on non-zero sum games, later on in our article.

It is worth mentioning that the topic of security games is closely related to adversarial learning, althought the roles and objectives of the learner and adversary in a security game are slightly different from those in adversarial learning. In security games~\cite{Paruchuri08}, the learner is called the defender, whose  objective is to protect a set of targets from an adversary, referred to as an attacker. The problem facing the defender is to allocate protection resources within budget and operational constraints, to achieve a desired level of security. Security games have been applied to different real-life applications including airport security~\cite{Pita11}, wildlife protection~\cite{Fang17} and natural resource conservation~\cite{Ford16}. An excellent discussion on security games is available in~\cite{Tambe11}. 

\section{Game Theory based Adversarial Learning Techniques}
\label{sec_gt_adv_learning}
Adversarial learning is usually modeled as a $2$-player game between the learner and adversary. The learner's set of actions corresponds to selecting different hyper-parameters for its classifer, while the adversary's set of actions correspond to different strategies for changing valid data into adversarial data. For example, the adversary's actions could be to add different amounts of perturbation or noise to valid data, or, to remove certain features from valid data.  
The utilities for the learner and the adversary are defined in terms of their joint actions. 

\begin{table*}
\begin{center}
\hspace*{-0.3in}
\begin{tabular}{|c|c|c|c|}
\hline
{\bf References}&{\bf Initial Information about learner}&{\bf Adversary Attack Model}& {\bf Validation domain}\\
& {\bf with adversary} & & \\
\hline
\rowcolor{Gray}
\multicolumn{4}{|c|}{\bf Zero Sum Games}\\
\hline
\cite{Dalvi04} & Full information about learner's & Causative attacks & Spam filtering\\
			   & utility, cost and  classifier parameters & &\\
\hline
\cite{Lowd05,Nelson10,Li14,Vorobeychik14}& No information about learner's & Exploratory attacks by changing & Spam filtering\\
									& utility, costs and classifier parameters&values of future  input&\\
\hline						
\cite{Globerson06,Dekel10,Teo07}& No information about learner's  & Exploratory attacks by removing & Spam filtering\\
									&  utility, costs and classifier parameters&  features from future input&\\
\hline
\cite{Hardt16}					& With and without info. about prob.& Exploratory attacks by changing & Spam filtering\\
								&  distribution of input and ground truth  & values of future  input &\\
\hline
\cite{Goodfellow14}				& Full information about learner's & Exploratory attacks counterfeiting  & Image classification\\									
								& utility, cost and classifier params &valid input & \\
\hline
\rowcolor{Gray}
\multicolumn{4}{|c|}{\bf Sequential, Bayesian, Non-zero Sum Games}\\
\hline
\cite{Bruckner11,Groshans13,Bruckner12}& No information about learner's & Exploratory attacks by changing  & Spam filtering\\
\cite{Groshans15,Mei15,Bulo16}				& utility, costs and  classifier parameters		& values of future input&\\
\hline
\cite{Alfeld17}& No information about learner's	& Exploratory attacks on test& Stock prices \\ 
& 	utility, costs and  classifier parameters &  set only & \\
\hline
\cite{Zhou12, Zhou14,Dritsoula17}&  No information about learner's           &Exploratory attacks by mixing valid& Spam filtering\\
								& utility, costs and classifier parameters	&  and adversarial input, e.g., altering  &\cite{Zhou12,Zhou14} \\
								&                                            & all or part of input features &\\
\hline
\end{tabular}
\caption{Comparison of game theory-based adversarial learning techniques.}
\label{table_summary}
\end{center}
\end{table*}

In one of the earliest and seminal works on adversarial classification Dalvi {\em et al.}~\cite{Dalvi04} formulated adversarial classification as a $2$-player competitive game between the learner and the adversary called a classification game. The learner's prediction mechanism is a binary classifier while the adversary creates adversarial input by perturbing features from legitimate input. The game is asymmetric as the adversary is aware of the learner's classifier parameters, utilities and costs, but the learner is not aware whether an input is adversarial versus legitimate. Within this setting, the learner's utility is defined as its value from classifying input (misclassification yields negative value), minus a per-feature cost for including input features in the classification algorithm. Similarly, the adversary's utility is defined as its value from misclassification of an adversarial input by the learner (correct classification by learner yields negative value to adversary), minus its cost to generate the adversarial input from legitimate input. Both learner and adversary play a Nash equilibrium strategy. The problem is formulated as a constrained optimization problem and solved as a mixed integer linear program for the adversary that is then used by the learner to determine a robust classification strategy. The proposed strategy is validated using spam email data sets with different adversarial perturbation strategies and different learner misclassification penalties and shown to yield positive classifier utilties, implying low misclassfication rates of adversarial input by the learner.  

Following ~\cite{Dalvi04}, researchers have proposed different approaches to define these utilities depending on their learner and adversary behavior models. We categorize the solution techniques proposed in literature into two categories - those that use zero-sum games to model the interaction between the learner and adversary followed by a minimax-based, linear optimization solution to solve it, and, those that model the interaction as a non-zero sum game and use a Nash equilibrium-based, bi-level optimization, or related solution. Table ~\ref{table_summary} gives a list of the major game theory-based approaches for adversarial learning along different learner and adversary models and solution techniques. 

\subsection{Zero Sum Games: Constrained Optimization-based Solution Techniques}
\label{sec_gt_zero_sum_games}

Lowd and Meek~\cite{Lowd05}, extended Dalvi {\em et al.}'s model while relaxing the assumption that the adversary has full information about the learner's classification algorithm, utilities and costs. Although their main algorithm called Adversarial Classifier Reverse Engineering (ACRE) is not based on a game, we review it here as it has formed the basis for future game theory-based techniques for adversarial learning. In the adversarial learning problem considered by~\cite{Lowd05}, the adversary can discover information about the learner's classifier by sending a limited number of queries containing adversarial input. With this limited knowledge of the learner's classifier, the adversary's objective is to determine the input instance that incurs the lowest cost to modify into an adversarial input. The set of such input instances are called attack vectors. Using the ACRE algorithm, the adversary can find attack vectors that can defeat the learner's classification algorithm when the input consists of either continuous or binary features. The algorithm is validated on spam email data sets when the learner uses either a Naive Bayes or a maximum entropy based classifier and shows that an adversary using ACRE can find instances within $17\%$ of the actual lowest cost instance while using a few thousand queries. In~\cite{Nelson10}, authors generalized ACRE to an $\epsilon$-instance minimal adversarial cost problem, where the learner's set of classifiers is expanded from linear classifier to more general, convex-inducing classifiers. Their solution searches over the adversarial cost space to determine the minimum set of adversarial examples, called the evasion set, that needs to be generated by the adversary to effect classification errors by the learner. Unlike~\cite{Lowd05}, avoiding the reverse engineering of the classifier allows their approach to handle classifiers that are difficult to reverse engineer. Recently, Li and Vorobeychik ~\cite{Li14}~\cite{Vorobeychik14} also extended the ACRE framework. Their proposed techniques include defining the cost between instances using equivalence class-based cost functions and solving the optimization problem facing the adversary as a mixed integer linear program, and, using a concept called moving target defense~\cite{Jajodia12} where the learner employs randomization over multiple classifiers instead of tuning parameters of a single classifier to make its prediction robust against adversarial attacks. 

Another parallel, but related line of research, considers the adversary's behavior from a slightly different approach, where the adversary generates adversarial data by selecting, removing or corrupting features from the input data set. Globerson {\em et al.}~\cite{Globerson06} describe such a setting where the adversary can remove multiple or single features from input and the learner's objective is to determine an optimal set of feature weights for its classifier, that minmize a metric called the hinge loss. The problem is formulated as a minimax, zero sum game and represented as a constrained optimization problem. The proposed algorithm was verified with adversarial data and shown to give lower error rates than a support vector machine classifier for handwritten digit classification and spam filtering. Globerson's model of selective feature removal by the adversary was extended in~\cite{Dekel10} with two variants of the learner. When the learner is able to train a classifier using training data, the problem is solved as a linear program. On the other hand, when the learner does not have access to training data, the learning task becomes an online learning problem. In this case, the learner's classifier is determined using a neural network, perceptron algorithm~\cite{Rosenblatt58} followed by a batching technique to model the online classifier as a statistical learning algorithm while making statistical guarantees about the classifier's performance. Experiments comparing the authors' algorithm with those in~\cite{Globerson06,Teo07} while using the same data sets, show that their proposed technique improves classification accuracy over the compared techniques.

In~\cite{Hardt16}, the authors consider adversarial learning games from the perspective of information revelation by the learner and adversary. They consider two variants of the learner: (i) when the learner has information about adversary's cost function, ground truth and input distribution, and (ii) when the learner knows only adversary's cost function, but not ground truth or input distribution. Similarly, two types of adversary are considered: one that knows all parameters that the learner knows plus its own adversarial example generation function, and, another, that knows only its own cost function and adversarial example generation function.  Within these settings, the learner's objective is to determine a {\em strategy robust algorithm} - one that selects a classifier  that maximizes the probability of the classifier's output corresponding to the ground truth for possibly adversarial examples. On the other hand, the adversary tries to create adversarial examples that maximize its utility  given by the difference between its benefit from the learner's classifier output for that example and its cost to generate the example. Their results theoretically analyze the running time and sample complexity of the learner for different types of adversary functions called separable and non-separable. 

As mentioned earlier, most existing literature on adversarial learning games assume a sequential-move game with the learner as the leader. However, few researchers have analyzed adversarial learning where the learner knows the adversary's strategy to generate adversarial data, but the adversary does not have information about the learner's classifier. This makes the adversary the leader and the learner the follower~\cite{Kantarcioglu11,Liu09}. Liu and Chawla~\cite{Liu09} consider such a setting where the adversary tries to generate adversarial data that results in moving the learner's classification boundary between spam versus non-spam email. The learner uses a genetic algorithm to search classifier parameters that reduce its classification error in response to these adversarial attacks. This work was extended by the authors to calculate the Nash equilibrium of a constant-sum game efficiently with reduced computation~\cite{Liu12}, and, recently, with the learner using a deep convolutional neural network as its classifier~\cite{Chivukula17}.  Nevertheless, several authors~\cite{Huang11,Bruckner11} justify the learner as the leader while observing that most real-life classifier-based systems like email spam filters and network traffic filters publish their classifier algorithms publicly. 

In contrast to sequential move games, simultaneous move games have been used less often to model interaction between an adversary and the learner. Recently, Schuurmans and Zinkevich~\cite{Schuurmans16} have addressed the problem of training a deep neural network as solving for a Nash equilibrium in a repeated, zero-sum game between two players, called protagonist and antagonist. The antagonist's objective is to determine a set of parameters that reduces the loss function during training, while the protagonist tries to select weight values of the edges in the deep network such that its utility is the negative of the antagonist's utility. Additional players called zannis are placed at the input and hidden layer nodes to select those nodes' parameter values. Iterations used to adjust weights of the neural network's edges in a conventional supervised training algorithm are modeled as repeated plays of the game aimed at converging to the Nash equilibrium and implemented using two algorithms called exponentiated weight and regret matching. Although not directly related to an adversarial setting where the adversary generates adversarial data to misguide the learner's classification, the deep learning game provides an interesting direction that is amenable to be extended to an adversarial setting.

\subsection{Non-zero Sum Games: Bayes-Nash Equilibrium and Related Solution Techniques}
\label{sec_nonzerosumgames}
Non-zero sum games have been proposed as a more realistic model for the interactions between the learner and adversary in adversarial learning~\cite{Bruckner11,Groshans13,Groshans15,Mei15,Alfeld17} because the loss in utility of learner might not exactly equal to the gain in utility of the adversary, and vice-versa. In a non-zero sum adversarial learning game, the learner calculates its loss in utility as proportional to the number of data instances on which it made a classification error because the data was modified by the adversary. However, a wrinkle to this approach is that the learner does not know whether the adversary had indeed modified the data to make it commit a classification mistake. For example, consider a learner in an automated email spam filter that has a spam identification rule as: 'if there are more than three misspelled words in an email text, then the email is spam'. Suppose that this learner receives an email that has five misspelled words and classifies it as spam following its spam identification rule. However, the learner does not know if the misspelled words were generated by an adversary, or, whether, they were genuine typographic errors made by a human. To address this problem, the learner tries to estimate probabilistically, if an instance it classifies was generated by an adversary or not, and, weighs its classification error by this probability. Correspondingly, the adversary also calculates its loss in utility depending on whether its adversarially generated instance was able to fool the classifier or not. Using this framework, Bruckner {\em et al.}~\cite{Bruckner11} proposed a non-zero sum game to model adversarial learning. The pairs of utilties of the learner and the adversary form a probabilistic version of the normal form game called a Bayesian game. The strategies adopted by the players in this Bayesian game, for example, the classifier hyperparameters selected by the learner and the perturbation strategy to modify legitimate data selected by the adversary, is calculated using the Bayes-Nash equilibrium.

Like the ACRE algorithm~\cite{Lowd05}, Bruckner's model~\cite{Bruckner11} has also been extended from different aspects in future research. In ~\cite{Mei15}, authors considered attacks on the training set in adversarial learning within a problem called machine teaching. Here, the adversary takes the role of a teacher while the learner takes on the role of a student whose objective is to learn a concept from data provided by the teacher. The objective of the teacher is to make causative attacks discussed in the background section of our article, so that it can coerce the learner towards learning a concept that it desires. Like~\cite{Bruckner11}, the problem is modeled as a bi-level optimization problem and solved by relaxing it to linear optimization using Karush-Kuhn Tucker(KKT) conditions. Subsequently, Alfeld {\em et al.}~\cite{Alfeld17} extended this framework to test time attacks on data. Bruckner's model~\cite{Bruckner11} has also been generalized in~\cite{Bulo16} using randomized prediction games where the learner's prediction algorithm randomizes over classifiers with different weight parameters while the adversary randomizes over its adversarial vectors. 

Another direction of adversarial learning investigated by Zhou {\em et al.}~\cite{Zhou12} is along the lines of Globerson's model of selective feature removal~\cite{Globerson06}. Here, the authors consider two adversarial attack models called full range attacks and restrained attacks. In a full range attack, the adversary can perturb any fraction of the maximum range of a feature. On the other hand, in a restrained range attack, the adversary can perturb only a fraction of the difference between its intended value and the actual value of a feature. The learner's prediction mechanism to counter these attacks is an SVM-based classifier that minimizes hinge loss. Extending this work towards more robust learner, the authors proposed a mixture of Bayesian experts approach~\cite{Zhou14} as the learner's prediction mechanism. 

An adversarial learning game called a classification game in~\cite{Dritsoula17} considers a practical adversarial strategy used by adversaries like spammers. Spammers might sometimes behave non-maliciously and not generate adversarial data to misguide the learner. This could result in false alarms by the learner if it incorrectly identifies the adversary as malicious when it is not. To account for this, the learner maintains a probability of the adversary being malicious versus non-malicious. The game is non-zero sum as the learner's utility includes the negative of the adversary's utility when it is a malicious, plus the learner's expected penalty from false alarms. Their work analyses the existence and uniqueness of Nash equilibrium for this classification game and proposes a constrained optimization solution, solved as linear program, to calculate the Nash equilibrium. Their model is validated by generating numeric values of learner and adversary costs, and size of strategy sets when data instances have either single or multiple features. Their results show that the learner utility decreases while attacker utility increases when either the cost of a single attack or the false alarm penalty increases.

\section{Learner Robustness via Adversarial Data Modeling}

The game theory-based adversarial learning techniques discussed thus far mainly focus on strategies that the learner could use to develop robustness against attacks from an adversary. Another approach to build the learner's defense mechanism, although not based on game theory, focusses on modeling the malicious data generated by the adversary so that the learner could understand the nature of adversarial attacks. Armed with information about the characteristics of adversarial attacks, the learner could then build appropriate defenses, such as train its classifier with the adversarial data, to improve robustness against the adversarial attacks. In this section, we provide an overview of three popular techniques for adversarial data generation that could be used in conjunction with adversarial learning. These techniques include adversarial data generation using perturbation techniques on valid examples, transferring adversarial examples across different learner models, and generative adversarial networks.

\subsection{Adversarial Data Generation via Perturbation} 
Before building the learner's defense mechanism against adversarial attacks, a first line of defense towards protecting against attacks is to understand how the adversary crafts those attacks. The topic of adversarial data generation seeks to address this issue by developing and analyzing different techniques that create synthetic, adversarial data, which could be used by potential adversaries. In most of these techniques, an adversarial example is constructed by adding a certain amount of noise or perturbation to a valid example. The main objective while creating an adversarial example is to perturb a valid example so that the perturbation is imperceptible to a human; in other words, the perturbed example appears to have the same label as a valid example to a human. However, when presented to a machine classifier, the same perturbed example would be assigned a different label than the valid example's label. For example, in a spam filtering scenario, when a perturbed example is created from a valid spam email message, the perturbed example would still appear to be a spam message to a human, but a spam filtering classifier would categorize it as non-spam, and, vice-versa. To achieve this property of imperceptibility to humans but deception for machine classifiers, the perturbation added to a valid example should take the perturbed example {\em just} across a decision boundary of the machine's classifier. Too little of a perturbation prevents the perturbed example from crossing the decision boundary - the perturbed example appears valid to the human, but does not fool the classifier either. On the other hand, too much perturbation takes the perturbed example far across the decision boundary, the classifier does not classify it correctly, but the excessively perturbed example appears as nonsense or rubbish~\cite{Goodfellow14-1} to a human who can easily discern it as an adversarially perturbed example. The main problem in adversarial data generation is then to determine this suitable amount of perturbation. 

In one of the earliest works in this direction, Biggio {\em et al.}~\cite{Biggio13} proposed a gradient descent technique that used the gradient of the discriminant function of the classifier along with the density function of the data to calculate a suitable amount of perturbation and generate a perturbed example. The proposed technique was validated to generate adversarial data from valid examples of handwritten digits and portable document format (PDF) text files while using different machine learning classifiers including linear classifiers, support vector machines and neural network classifiers. Optimization-based algorithms to determine the minimum amount of perturbation have been proposed in~\cite{Szegedy13}, ~\cite{Carlini16}. Goodfellow {\em et al.} in~\cite{Goodfellow14-1} made several fundamental contributions towards understanding properties of perturbations that create adversarial examples as well as properties of learner models that make them susceptible to adversarial examples. They proposed a fast, simple yet effective perturbation technique called Fast Gradient Sign Method (FGSM) to add perturbation proportional to the gradient of the cost function (e.g., loss function) to classify an example in a deep neural network. Their work also made valuable observations about perturbation techniques such as the direction of perturbation rather than amount of perturbation is more critical in creating adversarial examples, training a classifier with adversarial examples is akin to regularization of the classifier, and that a positive correlation exists between the degree to which a learning model can be optimized and its susceptibility to perturbation. Building upon these directions, researchers have proposed more refined perturbation techniques such as perturbing the label that has the lowest probability for the valid example in a single or multiple steps~\cite{Kurakin16-1}~\cite{Kurakin16-2}, perturbing an example's features that are most likely to change the classifier's output based on forward gradients~\cite{Papernot16}, universal peturbations to determine the minimal perturbation that will generate a certain fraction of adversarial examples guaranteed to result in misclassification when the examples are drawn from a given data distribution~\cite{Moosavi16} and neural networks called adversarial transformation networks that are trained to create adversarial examples~\cite{Baluja18}. Most of the aforementioned techniques have been proposed for generating adversarial data of handwritten digits or images. In contrast, Sethi and Kantardzic described methods to generate adversarial text data. Adversarial examples called attack data are constructed from probing the classifier with randomly perturbed text and retaining the perturbations that are successful in fooling the classifier, when the number of probes that the adversary can make is limited. Without limits on the number of probes, the adversary can reverse engineer~\cite{Lowd05} the classifier to create more precise advesarial examples that are able to fool the classifier more often~\cite{Sethi18}. In~\cite{Carlini18}, authors have recently proposed methods for generating adversarial audio data for misleading speech-to-text machine classifiers. In general, the topic of adversarial data generation techniques to gain better insights into how the adversary can deceive the learner's classifier with malicious data is still an open research problem that requires meticulous analysis of data perturbation techniques in conjunction with the characteristics of the model used by the learner for classification.

\subsection{Transferring Adversarial Examples}
The adversarial data generation techniques discussed in the previous section require the virtual adversary to utilize the model used by the learner to classify examples so that the virtual adversary can determine whether the adversarial examples it generates are able to deceive the learner's classifier. Because the virtual adversary cannot gain access to the learner's model, the adversary usually resorts to reconstructing the learner's model via probing - sending valid and adversarial examples to the learner's model of the classifier and observing the output label assigned by the classifier. Each probe incurs a cost for the adversary because the adversary has to expend resources to acquire valid examples and perturb them, plus the learner could limit the number of examples the adversary could send. To reduce costs, it would be beneficial for the adversary if it could generate adversarial examples to fool a classifier while utilizing a certain learner model, and then reuse those same adversarial examples to fool multiple, different classifiers. This technique of sending adversarial example generated using one learner model to a different model for classification is called {\em transferring} examples. The transfer problem is very relevant in the context of cybersecurity because it affords adversaries a low-cost technique to attack diverse machine learning-based classifiers such as email spam filters, network intrusion detection systems, identity authentication systems, etc., presumably at different locations while generating only one set of adversarial data.

As in the case of adversarial data generation, research in transferring adversarial examples has focused mainly on what characteristics of learner models of classifiers favor transferring adversarial examples across the models. In~\cite{Goodfellow14-1}, authors identified that when the weight vectors of two neural network-based learning models are aligned with each other, adversarial examples generated using one model could be transferred to the other. The transferability of adversarially generated image data across different learning-based models of image classifiers was investigated in~\cite{Liu16} to discover that the alignment of the decision boundaries of different models favors transferability of adversarial examples across models. Recently, Tramer {\em et al.}~\cite{Tramer17-2} investigated the dimensionality of adversarial subspaces as a means to determine the transferability of adversarial examples. They concluded that the subspace of adversarial examples has a large dimension (about $25$) and adversarial examples are transferable across two learner models when there is significant overlap in the subspace of the adversarial examples generated using the two models. Based on these findings the same authors proposed a technique called ensemble adversarial training~\cite{Tramer17-1} where perturbations generated using one learner model are transferred or informed to another learner model to make the latter model more robust to adversarial examples. Like adversarial data generation, transferability of adversarial examples between different learner models is also an open problem whose investigation could lead to more robust adversarial learning techniques.

\subsection{Generative Adversarial Networks}
Generative Adversarial Networks (GANs)~\cite{Goodfellow14} have recently been proposed as a game theory-based techniques to simultaneously generate perturbed examples from an adversary and then use those adversarial examples to train the learner's model. In a GAN, the learner uses a function called the discriminator (usually a classifier) while the adversary's function is called the generator. The interaction between the discriminator and the generator is modeled as a zero-sum game, and both the discriminator and generator iteratively solve a minimax optimization function. The optimization is done iteratively over chunks or batches of data and implements a gradient descent over the respective loss functions of the discriminator and generator. Experimental results of this approach show that both discriminator and generator are able to continuously adapt their prediction and data corruption mechanisms respectively.  Rather than improving the robustness of the learner to adversarial examples, the main contribution of GANs has been to demonstrate that the adversary (generator) can create very convincing adversarial or counterfeit examples, e.g., images of animals, human faces, traffic signs, etc. that are not distinguishable from a real image by the human eye, but can cause the discriminator to output an incorrect classification. For example, a picture of a cat could be modified by the generator in a way that is imperceptible to the human eye but causes the discriminator to mis-label it as an airplane. More generally, the GAN techniques have shown for the first time that data labeling, which is largely a supervised learning task utilizing labeled training data from humans, can also be implemented as an unsupervised learning task by exploiting the adversarial attacks of the generator. GANs have been used in several applications including image, audio and video generation, computer vision tasks such as image and video labeling~\cite{Vondrick16,Reed16}. Improvements of the basic GAN including Wasserstein GAN~\cite{Arjovsky17}, SeqGAN~\cite{Yu17}, StackGAN~\cite{Zhang16}, and EnergyGAN~\cite{Zhou16} have also been proposed. Defense mechanisms of the discriminator in a GAN include adversarial training, defensive distillation and gradient masking~\cite{Papernot17}, as well as statistical data analysis methods~\cite{Grosse17}, although many of these techniques have been shown to be vulnerable more recently~\cite{Carlini16},~\cite{He17}. Similar to perturbation-based adversarial data generation, techniques developed for GANs to generate adversarial examples could also be used towards developing stronger adversarial learning mechanisms.

\section{Open Problems and Further Directions}
\label{sec_future}
As discussed above, game theory offers a convenient means of modeling learner and adversary behaviors in adversarial learning. However, with recent developments in game theory based behavior modeling using repeated, evolutionary games and machine learning using deep networks, there are certain directions in which these models could be improved to make adversarial learning more robust against real-life adversarial threats. We identify some of these potential directions for future research below:

{\bf Richer models of learner and adversary behavior.} A shortcoming of many existing game theory based models of adversarial learning is that the learner uses only one-time interaction history with the adversary to build and update its model of the adversary’s behavior of generating adversarial data instances. We envisage that a more complex model, that considers the history of interactions between the learner and adversary can enable the learner to make more accurate decisions of the adversary's behavior and adapt its classifier accordingly. Repeated games with Bayesian learning provide a theoretical foundation for building such a history-based model of the adversary by the learner. Investigating this direction, while developing fast, heuristics-based algorithms that can guarantee accurate prediction of the adversary’s behavior within quantifiable bounds, would lead towards more accurate and robust models adversarial learning. Below, we identify some specific directions and open problems in adversarial learning.

{\bf Bounded life-force of adversary and learner.} Most of the game theory-based adversarial learning models discussed above assume that the adversary has unlimited resources (e.g., access to valid data, Internet access), and, budget to craft adversarial examples. Taken together, these assets could be considered as a {\em life force} of the adversary. However, in real-life, Internet adversaries such as spammers, usually have limited life force within which they attempt to maximize the impact of the adversarial data they generate. Recently, researchers have started exploring this direction by considering a bounded feature adversary~\cite{Park17} that is limited in the extent of change it can effect on features in valid data and on the number of queries it can make to the learner~\cite{Globerson06,Hardt16}. An interesting and practical future direction that has been less investigated is the effect diminishing life force on the strategy of the adversary. For example, an adversary that perceives very little remaining life force would adopt aggressive strategies to maximize its harm on the learner. Going a level deeper, the learner could also build a model of the adversary's life force by analyzing the adversary's attacks and strategize its defense mechanism to minimize harm. Repeated game frameworks with that incorporate life force-based strategizing offer an suitable direction to investigate this problem. 

{\bf Diminishing value of shared resources (data set).} Yet another limitation of existing adversarial learning models is that the value of shared resource, e.g., email data, is considered to be immutable while being subject to adversarial attacks. In reality, due to non-zero error rates of the learner's classifer, a small, but non-negligible amount of causative and exploratory attacks get past the classifier, giving rise to a compromised data set and reduce reliability of the classifier. It would make sense to investigate techniques that incorporate diminished value of data in training and test, and its effect on the classifier confidence, in the defense mechanism used by the learner. Once again, the adversary could simultaneously attempt to model the value of the data and the classifier confidence, and incorporate these metrics into the adversarial data generation strategy. 

{\bf Tactical defender and attacker strategies.} As discussed earlier, game theory based adversarial learning models use learner and adversary utility values to parameterize strategies. In real-life, adversaries like spammers of Website attackers use tactical strategies for their attacks. Example learner strategies could include guns or butter, growing soft, strict justice, layered defense, while the attacker could strategize with low but slow attacks, surprise attack, David and Goliath-type attack, suicide attack. Techniques from behavioral game theory provide a suitable framework for modeling such tactical strategies and bringing adversarial learning research closer to practical attacks.

{\bf Deeper behavior modeling by adversary and learner}. Most existing techniques for adversarial learning are based on a sequential game where the adversary has information about the classifier used by the learner, although it does not have information about the parameters of the classifier. An interesting and practical direction worthy of investigation is a game theory-based, adversarial learning model where the adversary has only partial, possibly inaccurate information about the learner’s classifier. This setting would be relevant in most real-life settings, as the classifier used by the learner is usually proprietary information that is confidential to the learner. Similarly, moving beyond Nash equlibrium, solutions like the price of anarchy, and regret minimization~\cite{LeytonBrown09} could provide faster means of calculating strategies by the learner and adversary. Related to this direction, game theory models could be made more informed by incorporating modeling and reasoning costs such as cost to solve for Nash equilibrium, cost to maintain game play history and build opponent models from the history. Similarly, expenses incurred by the adversary to get access to resources like legitimate email data sets and to the learner's classifier could be modeled as a reward it gets, that is proportional to its success in compromising the learner.

{\bf Robust classification with sparse data.}  Supervised learning algorithms, including support vector machines, regression learning, etc., that are used to build the learner’s classifier in adversarial learning rely heavily on large, information rich, training sets to predict correctly. In many instances, these algorithms suffer from low accuracy if the data used in training is sparse or does not contain all possible feature instances. To mitigate this problem, a technique called domain adaptation or transfer learning has been proposed in literature. However, to the best of our knowledge, transfer learning has not been investigated in the context of adversarial learning. Using transfer learning would be very relevant in adversarial learning with sparse training data. For example, the mapping determined by the classifier from input instances to class labels for helpful vs. not-helpful movie reviews in a movie reviews data set (e.g., IMDB data set), could be re-used, after suitable transformations, for classifying Internet clients as malicious vs. non-malicious from sparse, server log data. The critical problem here is to find correspondences between datasets between the source and target domains, and then suitably adapt the mapping learned in the source domain to the target domain.

\section{Conclusions}
In this paper, we have provided a systematic classification of adversarial learning techniques using game theoretic frameworks. While adversarial learning has been researched for over a decade, recent advances in machine learning, especially with deep networks, could be used to enhance existing game theory techniques for deeper learner and adversary behavior modeling, as well as, to compute more efficient and robust action selection strategies by the learner. We have identified several open problems and challenges for future research along these direcitons. With the recent, phenomenal growth of machine learning-based intelligent systems, we believe that addressing these challenges will advance real-life, classifier-based learning systems like email spam classifiers, social network sentiment analysis tools, and image and sensor data recognition systems on autonomous vehicles towards becoming more robust and reliable for seamless human use.

\bibliographystyle{abbrv}
\bibliography{refs}

\begin{thebibliography}{10}

\bibitem{Alfeld17}
S.~Alfeld, X.~Zhu, and P.~Barford.
\newblock Explicit defense actions against test-set attacks.
\newblock In {\em Proceedings of the Thirty-First {AAAI} Conference on
  Artificial Intelligence, February 4-9, 2017, San Francisco, California,
  {USA.}}, pages 1274--1280, 2017.

\bibitem{Arjovsky17}
M.~Arjovsky, S.~Chintala, and L.~Bottou.
\newblock Wasserstein generative adversarial networks.
\newblock In {\em International Conference on Machine Learning}, pages
  214--223, 2017.

\bibitem{Baluja18}
S.~Baluja and I.~Fischer.
\newblock Learning to attack: Adversarial transformation networks.
\newblock In {\em Proceedings of the Thirty-Second {AAAI} Conference on
  Artificial Intelligence, New Orleans, Louisiana, USA, February 2-7, 2018},
  2018.

\bibitem{Biggio13}
B.~Biggio, I.~Corona, D.~Maiorca, B.~Nelson, N.~{\v{S}}rndi{\'c}, P.~Laskov,
  G.~Giacinto, and F.~Roli.
\newblock Evasion attacks against machine learning at test time.
\newblock In {\em Joint European conference on machine learning and knowledge
  discovery in databases}, pages 387--402. Springer, 2013.

\bibitem{Bruckner12}
M.~Br\"{u}ckner, C.~Kanzow, and T.~Scheffer.
\newblock Static prediction games for adversarial learning problems.
\newblock {\em J. Mach. Learn. Res.}, 13(1):2617--2654, Sept. 2012.

\bibitem{Bruckner11}
M.~Bruckner and T.~Scheffer.
\newblock Stackelberg games for adversarial prediction problems.
\newblock In {\em Proceedings of the 17th ACM SIGKDD International Conference
  on Knowledge Discovery and Data Mining}, KDD '11, pages 547--555, 2011.

\bibitem{Bulo16}
S.~R. Bulò, B.~Biggio, I.~Pillai, M.~Pelillo, and F.~Roli.
\newblock Randomized prediction games for adversarial machine learning.
\newblock {\em IEEE Transactions on Neural Networks and Learning Systems},
  pages 1--13, 2016.

\bibitem{Carlini16}
N.~Carlini and D.~A. Wagner.
\newblock Towards evaluating the robustness of neural networks.
\newblock {\em CoRR}, abs/1608.04644, 2016.

\bibitem{Carlini18}
N.~Carlini and D.~A. Wagner.
\newblock Audio adversarial examples: Targeted attacks on speech-to-text.
\newblock {\em CoRR}, abs/1801.01944, 2018.

\bibitem{Chivukula17}
A.~S. Chivukula and W.~Liu.
\newblock Adversarial learning games with deep learning models.
\newblock In {\em Neural Networks (IJCNN), 2017 International Joint Conference
  on}, pages 2758--2767. IEEE, 2017.

\bibitem{Dalvi04}
N.~Dalvi, P.~Domingos, Mausam, S.~Sanghai, and D.~Verma.
\newblock Adversarial classification.
\newblock In {\em Proceedings of the Tenth ACM SIGKDD International Conference
  on Knowledge Discovery and Data Mining}, KDD '04, pages 99--108, 2004.

\bibitem{Dekel10}
O.~Dekel, O.~Shamir, and L.~Xiao.
\newblock Learning to classify with missing and corrupted features.
\newblock {\em Machine Learning}, 81(2):149--178, 2010.

\bibitem{Dritsoula17}
L.~Dritsoula, P.~Loiseau, and J.~Musacchio.
\newblock A game-theoretic analysis of adversarial classification.
\newblock {\em IEEE Transactions on Information Forensics and Security},
  PP(99):1--1, 2017.

\bibitem{Fang17}
F.~Fang, T.~H. Nguyen, R.~Pickles, W.~Y. Lam, G.~R. Clements, B.~An, A.~Singh,
  B.~C. Schwedock, M.~Tambe, and A.~Lemieux.
\newblock {PAWS} - {A} deployed game-theoretic application to combat poaching.
\newblock {\em {AI} Magazine}, 38(1):23--36, 2017.

\bibitem{Ford16}
B.~J. Ford, M.~Brown, A.~Yadav, A.~Singh, A.~Sinha, B.~Srivastava,
  C.~Kiekintveld, and M.~Tambe.
\newblock Protecting the {NECTAR} of the ganga river through game-theoretic
  factory inspections.
\newblock In {\em Advances in Practical Applications of Scalable Multi-agent
  Systems. The {PAAMS} Collection - 14th International Conference, {PAAMS}
  2016, Sevilla, Spain, June 1-3, 2016, Proceedings}, pages 97--108, 2016.

\bibitem{Fudenberg91}
D.~Fudenberg and J.~Tirole.
\newblock {\em Game Theory}.
\newblock MIT Press, 1991.

\bibitem{Globerson06}
A.~Globerson and S.~T. Roweis.
\newblock Nightmare at test time: robust learning by feature deletion.
\newblock In {\em Machine Learning, Proceedings of the Twenty-Third
  International Conference {(ICML} 2006), Pittsburgh, Pennsylvania, USA, June
  25-29, 2006}, pages 353--360, 2006.

\bibitem{Goodfellow14}
I.~J. Goodfellow, J.~Pouget{-}Abadie, M.~Mirza, B.~Xu, D.~Warde{-}Farley,
  S.~Ozair, A.~C. Courville, and Y.~Bengio.
\newblock Generative adversarial nets.
\newblock In {\em Advances in Neural Information Processing Systems 27: Annual
  Conference on Neural Information Processing Systems 2014, December 8-13 2014,
  Montreal, Quebec, Canada}, pages 2672--2680, 2014.

\bibitem{Goodfellow14-1}
I.~J. Goodfellow, J.~Shlens, and C.~Szegedy.
\newblock Explaining and harnessing adversarial examples.
\newblock {\em CoRR}, abs/1412.6572, 2014.

\bibitem{Grosse17}
K.~Grosse, P.~Manoharan, N.~Papernot, M.~Backes, and P.~D. McDaniel.
\newblock On the (statistical) detection of adversarial examples.
\newblock {\em CoRR}, abs/1702.06280, 2017.

\bibitem{Groshans13}
M.~Grosshans, C.~Sawade, M.~Bruckner, and T.~Scheffer.
\newblock Bayesian games for adversarial regression problems.
\newblock In {\em Proceedings of the 30th International Conference on
  International Conference on Machine Learning - Volume 28}, ICML'13, pages
  III--55--III--63, 2013.

\bibitem{Groshans15}
M.~Grosshans and T.~Scheffer.
\newblock Solving prediction games with parallel batch gradient descent.
\newblock In {\em Proceedings of the 2015th European Conference on Machine
  Learning and Knowledge Discovery in Databases - Volume Part I}, ECMLPKDD'15,
  pages 152--167, 2015.

\bibitem{Hardt16}
M.~Hardt, N.~Megiddo, C.~Papadimitriou, and M.~Wootters.
\newblock Strategic classification.
\newblock In {\em Proceedings of the 2016 ACM Conference on Innovations in
  Theoretical Computer Science}, ITCS '16, pages 111--122, 2016.

\bibitem{Harsanyi68}
J.~Harsanyi.
\newblock Games with incomplete information played by `bayesian' players, part
  iii. the basic probability distribution of the game.
\newblock {\em Management Science}, 14(7):486--502, 1968.

\bibitem{He17}
W.~He, J.~Wei, X.~Chen, N.~Carlini, and D.~Song.
\newblock Adversarial example defense: Ensembles of weak defenses are not
  strong.
\newblock In {\em 11th {USENIX} Workshop on Offensive Technologies, {WOOT}
  2017, Vancouver, BC, Canada, August 14-15, 2017.}, 2017.

\bibitem{Huang11}
L.~Huang, A.~D. Joseph, B.~Nelson, B.~I. Rubinstein, and J.~D. Tygar.
\newblock Adversarial machine learning.
\newblock In {\em Proceedings of the 4th ACM Workshop on Security and
  Artificial Intelligence}, AISec '11, pages 43--58, 2011.

\bibitem{Jajodia12}
S.~Jajodia, A.~K. Ghosh, V.~S. Subrahmanian, V.~Swarup, C.~Wang, and X.~S.
  Wang.
\newblock {\em Moving Target Defense II: Application of Game Theory and
  Adversarial Modeling}.
\newblock Springer Publishing Company, Incorporated, 2012.

\bibitem{Kantarcioglu11}
M.~Kantarcioglu, B.~Xi, and C.~Clifton.
\newblock Classifier evaluation and attribute selection against active
  adversaries.
\newblock {\em Data Min. Knowl. Discov.}, 22(1-2):291--335, 2011.

\bibitem{Kurakin16-1}
A.~Kurakin, I.~J. Goodfellow, and S.~Bengio.
\newblock Adversarial examples in the physical world.
\newblock {\em CoRR}, abs/1607.02533, 2016.

\bibitem{Kurakin16-2}
A.~Kurakin, I.~J. Goodfellow, and S.~Bengio.
\newblock Adversarial machine learning at scale.
\newblock {\em CoRR}, abs/1611.01236, 2016.

\bibitem{Li14}
B.~Li and Y.~Vorobeychik.
\newblock Feature cross-substitution in adversarial classification.
\newblock In {\em Advances in Neural Information Processing Systems 27: Annual
  Conference on Neural Information Processing Systems 2014, December 8-13 2014,
  Montreal, Quebec, Canada}, pages 2087--2095, 2014.

\bibitem{Liu09}
W.~Liu and S.~Chawla.
\newblock {\em A game theoretical model for adversarial learning}, pages
  25--30.
\newblock 2009.

\bibitem{Liu12}
W.~Liu, S.~Chawla, J.~Bailey, C.~Leckie, and K.~Ramamohanarao.
\newblock An efficient adversarial learning strategy for constructing robust
  classification boundaries.
\newblock In {\em Australasian Joint Conference on Artificial Intelligence},
  pages 649--660. Springer, 2012.

\bibitem{Liu16}
Y.~Liu, X.~Chen, C.~Liu, and D.~Song.
\newblock Delving into transferable adversarial examples and black-box attacks.
\newblock {\em CoRR}, abs/1611.02770, 2016.

\bibitem{Lowd05}
D.~Lowd and C.~Meek.
\newblock Adversarial learning.
\newblock In {\em Proceedings of the Eleventh ACM SIGKDD International
  Conference on Knowledge Discovery in Data Mining}, KDD '05, pages 641--647,
  2005.

\bibitem{Mei15}
S.~Mei and X.~Zhu.
\newblock Using machine teaching to identify optimal training-set attacks on
  machine learners.
\newblock In {\em Proceedings of the Twenty-Ninth AAAI Conference on Artificial
  Intelligence}, AAAI'15, pages 2871--2877. AAAI Press, 2015.

\bibitem{Moosavi16}
S.~Moosavi{-}Dezfooli, A.~Fawzi, O.~Fawzi, and P.~Frossard.
\newblock Universal adversarial perturbations.
\newblock {\em CoRR}, abs/1610.08401, 2016.

\bibitem{Myerson97}
R.~B. Myerson.
\newblock {\em Game Theory: Analysis of Conflict}.
\newblock Harvard University Press, 1997.

\bibitem{Nelson10}
B.~Nelson, B.~I.~P. Rubinstein, L.~Huang, A.~D. Joseph, S.~Lau, S.~J. Lee,
  S.~Rao, A.~Tran, and J.~D. Tygar.
\newblock Near-optimal evasion of convex-inducing classifiers.
\newblock In {\em Proceedings of the Thirteenth International Conference on
  Artificial Intelligence and Statistics, {AISTATS} 2010, Chia Laguna Resort,
  Sardinia, Italy, May 13-15, 2010}, pages 549--556, 2010.

\bibitem{Papernot17}
N.~Papernot, P.~McDaniel, I.~Goodfellow, S.~Jha, Z.~B. Celik, and A.~Swami.
\newblock Practical black-box attacks against machine learning.
\newblock In {\em Proceedings of the 2017 ACM on Asia Conference on Computer
  and Communications Security}, ASIA CCS '17, pages 506--519, 2017.

\bibitem{Papernot16}
N.~Papernot, P.~McDaniel, S.~Jha, M.~Fredrikson, Z.~B. Celik, and A.~Swami.
\newblock The limitations of deep learning in adversarial settings.
\newblock In {\em Security and Privacy (EuroS\&P), 2016 IEEE European Symposium
  on}, pages 372--387. IEEE, 2016.

\bibitem{Park17}
S.~Park, J.~Weimer, and I.~Lee.
\newblock Resilient linear classification: An approach to deal with attacks on
  training data.
\newblock In {\em Proceedings of the 8th International Conference on
  Cyber-Physical Systems}, ICCPS '17, pages 155--164, 2017.

\bibitem{Paruchuri08}
P.~Paruchuri, J.~P. Pearce, J.~Marecki, M.~Tambe, F.~Ord{\'{o}}{\~{n}}ez, and
  S.~Kraus.
\newblock Playing games for security: an efficient exact algorithm for solving
  bayesian stackelberg games.
\newblock In {\em 7th International Joint Conference on Autonomous Agents and
  Multiagent Systems {(AAMAS} 2008), Estoril, Portugal, May 12-16, 2008, Volume
  2}, pages 895--902, 2008.

\bibitem{Pita11}
J.~Pita, M.~Tambe, C.~Kiekintveld, S.~Cullen, and E.~Steigerwald.
\newblock {GUARDS} - innovative application of game theory for national airport
  security.
\newblock In {\em {IJCAI} 2011, Proceedings of the 22nd International Joint
  Conference on Artificial Intelligence, Barcelona, Catalonia, Spain, July
  16-22, 2011}, pages 2710--2715, 2011.

\bibitem{Reed16}
S.~Reed, Z.~Akata, X.~Yan, L.~Logeswaran, B.~Schiele, and H.~Lee.
\newblock Generative adversarial text to image synthesis.
\newblock {\em arXiv preprint arXiv:1605.05396}, 2016.

\bibitem{Rosenblatt58}
F.~Rosenblatt.
\newblock The perceptron: A probabilistic model for information storage and
  organization in the brain.
\newblock {\em Psychological Review}, pages 65--386, 1958.

\bibitem{Schuurmans16}
D.~Schuurmans and M.~A. Zinkevich.
\newblock Deep learning games.
\newblock In D.~D. Lee, M.~Sugiyama, U.~V. Luxburg, I.~Guyon, and R.~Garnett,
  editors, {\em Advances in Neural Information Processing Systems 29}, pages
  1678--1686. Curran Associates, Inc., 2016.

\bibitem{Sethi18}
T.~S. Sethi and M.~Kantardzic.
\newblock Data driven exploratory attacks on black box classifiers in
  adversarial domains.
\newblock {\em Neurocomputing}, 289:129--143, 2018.

\bibitem{LeytonBrown09}
Y.~Shoham and K.~Leyton{-}Brown.
\newblock {\em Multiagent Systems - Algorithmic, Game-Theoretic, and Logical
  Foundations}.
\newblock Cambridge University Press, 2009.

\bibitem{Szegedy13}
C.~Szegedy, W.~Zaremba, I.~Sutskever, J.~Bruna, D.~Erhan, I.~Goodfellow, and
  R.~Fergus.
\newblock Intriguing properties of neural networks.
\newblock {\em arXiv preprint arXiv:1312.6199}, 2013.

\bibitem{Tambe11}
M.~Tambe.
\newblock {\em Security and Game Theory: Algorithms, Deployed Systems, Lessons
  Learned}.
\newblock Cambridge University Press, New York, NY, USA, 1st edition, 2011.

\bibitem{Teo07}
C.~H. Teo, A.~Globerson, S.~T. Roweis, and A.~J. Smola.
\newblock Convex learning with invariances.
\newblock In {\em Advances in Neural Information Processing Systems 20,
  Proceedings of the Twenty-First Annual Conference on Neural Information
  Processing Systems, Vancouver, British Columbia, Canada, December 3-6, 2007},
  pages 1489--1496, 2007.

\bibitem{Tramer17-1}
F.~Tram{\`{e}}r, A.~Kurakin, N.~Papernot, D.~Boneh, and P.~D. McDaniel.
\newblock Ensemble adversarial training: Attacks and defenses.
\newblock {\em CoRR}, abs/1705.07204, 2017.

\bibitem{Tramer17-2}
F.~Tramèr, N.~Papernot, I.~Goodfellow, D.~Boneh, and P.~McDaniel.
\newblock The space of transferable adversarial examples.
\newblock {\em arXiv}, 2017.

\bibitem{Tygar11}
J.~D. Tygar.
\newblock Adversarial machine learning.
\newblock {\em {IEEE} Internet Computing}, 15(5):4--6, 2011.

\bibitem{Vondrick16}
C.~Vondrick, H.~Pirsiavash, and A.~Torralba.
\newblock Generating videos with scene dynamics.
\newblock In {\em Advances In Neural Information Processing Systems}, pages
  613--621, 2016.

\bibitem{Vorobeychik14}
Y.~Vorobeychik and B.~Li.
\newblock Optimal randomized classification in adversarial settings.
\newblock In {\em International conference on Autonomous Agents and Multi-Agent
  Systems, {AAMAS} '14, Paris, France, May 5-9, 2014}, pages 485--492, 2014.

\bibitem{Yu17}
L.~Yu, W.~Zhang, J.~Wang, and Y.~Yu.
\newblock Seqgan: Sequence generative adversarial nets with policy gradient.
\newblock In {\em AAAI}, pages 2852--2858, 2017.

\bibitem{Zhang16}
H.~Zhang, T.~Xu, H.~Li, S.~Zhang, X.~Huang, X.~Wang, and D.~N. Metaxas.
\newblock Stackgan: Text to photo-realistic image synthesis with stacked
  generative adversarial networks.
\newblock {\em CoRR}, abs/1612.03242, 2016.

\bibitem{Zhou14}
Y.~Zhou and M.~Kantarcioglu.
\newblock Adversarial learning with bayesian hierarchical mixtures of experts.
\newblock In {\em Proceedings of the 2014 {SIAM} International Conference on
  Data Mining, Philadelphia, Pennsylvania, USA, April 24-26, 2014}, pages
  929--937, 2014.

\bibitem{Zhou16}
Y.~Zhou and M.~Kantarcioglu.
\newblock Modeling adversarial learning as nested stackelberg games.
\newblock In {\em Advances in Knowledge Discovery and Data Mining - 20th
  Pacific-Asia Conference, {PAKDD} 2016, Auckland, New Zealand, April 19-22,
  2016, Proceedings, Part {II}}, pages 350--362, 2016.

\bibitem{Zhou12}
Y.~Zhou, M.~Kantarcioglu, B.~M. Thuraisingham, and B.~Xi.
\newblock Adversarial support vector machine learning.
\newblock In {\em The 18th {ACM} {SIGKDD} International Conference on Knowledge
  Discovery and Data Mining, {KDD} '12, Beijing, China, August 12-16, 2012},
  pages 1059--1067, 2012.

\end{thebibliography}

\end{document}